\documentclass[aps, prd, twocolumn, superscriptaddress, nofootinbib]{revtex4-1}
\usepackage{graphicx}
\usepackage{graphics}
\usepackage{dcolumn}
\usepackage{amssymb}
\usepackage{bm}
\usepackage{amsmath}
\usepackage{color}
\usepackage{enumerate}

\newcommand{\be}{\begin{equation}}
\newcommand{\ee}{\end{equation}}
\newcommand{\bea}{\begin{eqnarray}}
\newcommand{\eea}{\end{eqnarray}}

\begin{document}

\title{Kinematics of particles with quantum de Sitter symmetries}

\author{Leonardo Barcaroli}
\email{leonardo.barcaroli@roma1.infn.it}
\affiliation{Dipartimento di Fisica, Universit\`a "La Sapienza''
and Sez. Roma1 INFN, P.le A. Moro 2, 00185 Roma, Italy}
\author{Giulia Gubitosi}
\email{g.gubitosi@imperial.ac.uk}
\affiliation{Theoretical Physics, Blackett Laboratory, Imperial College, London SW7 2AZ, United Kingdom.}

\begin{abstract}
We present the first detailed study of the kinematics of free relativistic particles whose symmetries are described by a quantum deformation of the de Sitter algebra, known as $q$-de Sitter Hopf algebra. The quantum deformation parameter is a function of the Planck length $\ell$ and the de Sitter radius $H^{-1}$, such that when the Planck length vanishes, the algebra reduces to the de Sitter algebra, while when the de Sitter radius is sent to infinity one recovers the $\kappa$-Poincar\'e Hopf algebra. In the first limit the picture is that of a particle with trivial momentum space geometry moving on de Sitter spacetime, in the second one the picture is that of a particle with de Sitter momentum space geometry moving on Minkowski spacetime.
When both the Planck length and the inverse of the de Sitter radius are non-zero,  effects due to spacetime curvature and non-trivial momentum space geometry are both present and affect each other. The particles' motion is then described in a full phase space picture. We find that redshift effects that are usually associated to spacetime curvature become energy-dependent. Also, the energy dependence of particles' travel times that is usually associated to momentum space non-trivial properties is modified in a curvature-dependent way.
\end{abstract}

\maketitle

\section{Introduction}
Phenomenological models implementing relativistic Planck-scale-modified dispersion relations have gained considerable attention in the quantum gravity literature \cite{AmelinoCamelia:2000mn,AmelinoCamelia:2000ge}. They in fact describe effects that are quite generically expected in quantum gravity research \cite{AmelinoCamelia:2008qg,Mattingly:2005re} (and in particular have been explicitly shown to characterize 3D quantum gravity \cite{Freidel:2003sp,Freidel:2005bb}) without introducing preferred frames and so evading the strong constraints on Lorentz invariance violations. Modified dispersion relations could produce observable phenomenology in the context of astrophysics \cite{AmelinoCamelia:2009pg,AmelinoCamelia:2003ex, AmelinoCamelia:1997gz,Ackermann:2009aa,Amelino-Camelia:2013naa} and there is also increasing evidence that they could be relevant in the early universe \cite{Amelino-Camelia:2013tla,Amelino-Camelia:2013wha,Amelino-Camelia:2013gna,Amelino-Camelia:2015dqa,Arzano:2015gda,Gubitosi:2015osv,Magueijo:2008yk,Santos:2015sva, Awad:2013nxa}.   

Hopf algebras provide a consistent theoretical framework to describe the sort of deformations of spacetime symmetries required to introduce an invariant energy scale, and thus accommodate modifications of particles' dispersion relations, without spoiling the relativity principle. In this context, one of the most studied models is the one described by the $\kappa$-Poincar\'e  Hopf algebra \cite{Majid:1994cy,Lukierski:1991pn,Lukierski:1992dt,Lukierski:1993wxa}, a quantum deformation of the special-relativistic Poincar\'e group. $\kappa$-Poincar\'e symmetries have been shown to characterize the kinematics of particles living on a flat spacetime and non-trivial momentum space with a de Sitter geometry \cite{Gubitosi:2013rna, AmelinoCamelia:2011nt,KowalskiGlikman:2003we,KowalskiGlikman:2002ft} \footnote{Non-trivial momentum space geometry is a general feature of relativistic theories introducing an invariant energy scale  \cite{AmelinoCamelia:2011bm, AmelinoCamelia:2011pe, Amelino-Camelia:2013sba,Arzano:2014jua}. }.

Despite the fact that most of the research on relativistically compatible deformations of particles' kinematics focuses on cases where spacetime is flat, as mentioned before the  best opportunities for phenomenology are found in contexts where spacetime curvature should not be neglected.  Only very recently, after early attempts \cite{Marciano:2010gq, Mignemi:2008ek,Mignemi:2008fj, KowalskiGlikman:2004kp} that were however lacking a full understanding of the relative-locality effects produced by momentum space curvature \cite{AmelinoCamelia:2011bm, AmelinoCamelia:2011pe, Amelino-Camelia:2013sba}, there have been some proposals to coherently describe non-trivial momentum space properties alongside curvature of spacetime in a relativistic way. Some \cite{Barcaroli:2015xda,Amelino-Camelia:2014rga,Cianfrani:2014fia} focussed on finding an appropriate geometrical description of phase space. Others \cite{AmelinoCamelia:2012it,Rosati:2015pga} opted for a more phenomenological approach, aimed at building the appropriately modified algebra of symmetries, compatibly with the introduction of a curvature invariant besides the speed of light invariant and an energy scale invariant.  
Here we take a similar perspective as the one of these last studies, but working within the safe boundaries of Hopf algebras, which guarantee that not only the algebra but also the extra structures required for a relativistic theory (such as conservations laws) can be built in a consistent way. We focus on a quantum deformation of the de Sitter algebra (the algebra of isometries of the de Sitter spacetime), known as $q$-de Sitter \cite{0305-4470-26-21-019, Ballesteros:2004eu, Ballesteros:2013fh, Lukierski:1991ff}.  The  dimensionless quantum deformation parameter of the $q$-de Sitter algebra can be fixed as a function of the Planck length $\ell$ and the de Sitter radius $H^{-1}$, and we chose it so that when the Planck length vanishes, the algebra reduces to the de Sitter algebra, while when the de Sitter radius is sent to infinity one recovers the $\kappa$-Poincar\'e Hopf algebra. Studying the kinematics of particles' in these two limits corresponds to study the two complementary cases \cite{Amelino-Camelia:2013uya} in which either spacetime or momentum space have de Sitter geometry. The general case, with both $\ell$ and $H$ different from zero, allows to study the kinematics of particles' living on a phase space with de Sitter geometry both in the spacetime  and the momentum space sides.  As we will show, not only one recovers the effects expected in the two limiting cases, but the interplay between the non-trivial geometrical properties of the two parts of the phase space lead to novel effects \footnote{The $q$-de Sitter Hopf algebra was already considered from a phenomenological perspective in \cite{Marciano:2010gq}, where hints about the non-trivial interplay between curvature and Planck-scale effects were provided, but at the time there was no clear understanding of the way to properly handle kinematics in models with non-trivial momentum space geometry, and in particular the issue of relating observations made by different observers had not been clarified by the understanding of relative locality \cite{AmelinoCamelia:2011bm,AmelinoCamelia:2011pe}. 
}. 

The study of the kinematical properties of particles with $q$-de Sitter symmetries requires tools that were already successfully used for particles with $\kappa$-Poincar\'e symmetries and for particles living on a de Sitter spacetime with trivial momentum space \cite{Amelino-Camelia:2013uya}.
In section \ref{sec:dS} we review briefly this last case, in order to introduce the methodology and the notation in a context that is familiar to most readers. We introduce the de Sitter algebra, write down the action of finite translations and the evolution of the phase space coordinates. We then expose a derivation of the well-known redshift effect affecting particles traveling in such spacetime.  Section \ref{sec:kP} provides a similar analysis for the case of a particle moving on flat spacetime but with de Sitter geometry on momentum space, whose symmetries are described by the $\kappa$-Poincar\'e Hopf algebra in the bicrossproduct basis. Of course in this case there is no redshift effect, but a complementary effect \cite{Amelino-Camelia:2013uya} is present, which makes the travel time of particles between two observers to depend on the particles' energy. The full analysis of the kinematics described by the $q$-de Sitter algebra is done in Section \ref{sec:qdS}. We derive the action of finite translations on the phase space coordinates, write down the particles' worldlines in the full phase space and work out some possibly observable effects due to the interplay between curvature of spacetime and of momentum space. In particular, in subsection \ref{subs:qdSredshift} we observe that the amount of redshift undergone by a particle's energy during propagation is dependent on the initial energy of the particle besides the travel time. Moreover, in subsection \ref{subs:qdSlateshift} we show that the delay in travel time of particles with different energies, a feature characteristic of curved momentum space models, becomes dependent on the de Sitter radius parameter as well.

We work in $1+1$ dimensions and use a representation of phase space coordinates,  $x^{\mu}$ and $p_{\mu}$, $\mu=\{0,1\}$, with standard symplectic structure:
\bea
\{x^{\mu},x^{\nu}\}= & 0\,,\nonumber\\
\{x^{\mu},p_{\nu}\}= & -\delta_{\nu}^{\mu}\,,\label{eq:sympl}\\
\{p_{\mu},p_{\nu}\}= &\nonumber 0\,.
\eea

The action of finite spacetime translations (generated by the operators $\mathcal P_{0}$ and $\mathcal P_{1}$ and with translation parameters $a^{0}$ and $a^{1}$) on a phase space function $F(x^{\mu},p_{\nu})$ is in general found via the ordinary left action: 
\begin{equation}
\mathcal T_{\{a^{0},a^{1}\}} \triangleright F= \sum_{n=0}^{\infty} \frac{1}{n!} \underbrace{\{-a^{\mu}\mathcal P_{\mu},\{\dots,\{-a^{\mu}\mathcal P_{\mu}}_{n\,\text{times}}, F\}\dots\}\,,
\label{eq:FiniteTranslations}
\end{equation}
where in $1+1$ dimensions $a^{\mu}\mathcal P_{\mu}=a^{0}\mathcal P_{0}+a^{1}\mathcal P_{1}$.
\section{de Sitter spacetime}
\label{sec:dS}
This section reviews well-known facts about kinematics of particles on de Sitter spacetime, with a slightly different approach than the one most readers might be used to. The scope is to introduce notation and procedures that might look convoluted at this stage, but will become useful in the following sections, when dealing with the $\kappa$-Poincar\'e and $q$-de Sitter symmetries. We use comoving coordinates for spacetime and the corresponding dual coordinates for momentum space. The results reported here are derived in more detail in \cite{Amelino-Camelia:2013uya}.

\subsection{de Sitter algebra}

De Sitter spacetime is maximally symmetric, and as such it has three generators of global symmetry transformations, $\{\mathcal P_{0},\mathcal P_{1}, \mathcal N\}$, which are, respectively, the time translation, space translation and boost generators. Their algebra reads, at first order in the inverse of the de Sitter radius $H$:
\begin{eqnarray}
\{\mathcal P_{0},\mathcal P_{1}\}&= & H\,\mathcal P_{1}\,,\nonumber\\
{}\{\mathcal P_{0},\mathcal{N}\}&= &\mathcal P_{1}-H\,\mathcal{N} \,,\label{eq:dSalgebra}\\
{}\{\mathcal P_{1},\mathcal{N}\}&= &\mathcal P_{0}\,,\nonumber
\end{eqnarray}
and the Casimir of this algebra is:
\be
\mathcal C_{dS}= \mathcal P_{0}^{2}-\mathcal P_{1}^{2}+2 H \mathcal N \mathcal P_{1}.
\ee
Upon introducing the standard symplectic structure on the phase space coordinates $x^{\mu}$ and $p_{\mu}$, eq. (\ref{eq:sympl}), we can represent the generators as:
\bea
\mathcal P_{0}&=& p_{0}-H x^{1} p_{1}\,,\nonumber\\
\mathcal P_{1}&=& p_{1}\,,\label{eq:dSgeneratos}\\
\mathcal N&=&p_{1} \,x^{0}+p_{0}\, x^{1} -H \left( p_{1}(x^{0})^{2}+\frac{1}{2}p_{1}\,(x^{1})^{2}\right)\,,\nonumber
\eea
and the Casimir as:
\be
\mathcal C_{dS}=p_{0}^{2}-p_{1}^{2}+2 H p_{1}^{2} \,x^{0}\,.\label{eq:dSCasimir}
\ee

The action of finite spacetime translations on the phase space coordinates can be easily derived to be:
\bea
x^{0}_{B}&\equiv& \mathcal T_{\{a^{0},a^{1}\}} \triangleright x^{0}_{A} = x^{0}_{A}-a^{0}\,, \nonumber\\
x^{1}_{B}&\equiv& \mathcal T_{\{a^{0},a^{1}\}} \triangleright x^{1}_{A} = x^{1}_{A}(1+H a^{0})-a^{1}(1+\frac{1}{2} Ha^{0}) \,,\nonumber\\
p^{B}_{0}&\equiv& \mathcal T_{\{a^{0},a^{1}\}} \triangleright p^{A}_{0} = p^{A}_{0}
\,,\label{eq:dStranslations}\\
p^{B}_{1}&\equiv& \mathcal T_{\{a^{0},a^{1}\}} \triangleright p^{A}_{1} = p^{A}_{1}(1-H a^{0})
\,,\nonumber
\eea
where the indices $A,B$ indicate two observers linked by the spacetime translation  with translation parameters $\{a^{0}, a^{1}\}$ and we used the general prescription of eq. (\ref{eq:FiniteTranslations}). 

\subsection{Kinematics of massless particles in de Sitter spacetime and redshift}
The kinematics of a free massless particle moving on de Sitter spacetime is governed by the Hamilton equations, obtained using the Casimir (\ref{eq:dSCasimir}) as Hamiltonian:
\bea
\dot{x}^{0}&\equiv&\{\mathcal{C}_{dS},x^{0}\}=
 2p_{0}\,,\nonumber\\
\dot{x}^{1}&\equiv&\{\mathcal{C}_{dS},x^{1}\}= -2p_{1}(1-2H\, x^{0})\,,\nonumber\\
\dot{p}_{0}&\equiv&\{\mathcal{C}_{dS},{p}_{0}\}=   -2H\, p_{1}^{2}\,,\label{eq:dSHamilton}\\
\dot{p}_{1}&\equiv&\{\mathcal{C}_{dS},{p}_{1}\}= 0\,,\nonumber
\eea
where over-dots indicate derivatives with respect to the worldline's affine parameter $\tau$. Momenta $\{p_{0},p_{1}\}$ have to satisfy the mass-shell constraint  $\mathcal C_{dS}=0$ throughout their evolution along the particle's worldline:
\be
p_{0}=-p_{1}(1-H x^{0})\,.\label{eq:dSmasslessconstraint}
\ee
We have chosen the negative-sign solution to the mass-shell constraint in order to have positive coordinate velocity:
\be
v\equiv \frac{\dot{x}^{1}}{\dot{x}^{0}}= -\frac{p_{1}}{p_{0}}(1-2 H x^{0})=1-H x^{0}\,.
\ee
The particle's worldline can be found by integrating the coordinate velocity along the coordinate time $x^{0}$:
\bea
x^{1}-\bar x^{1}&\equiv&\int_{0}^{\tau}\dot x^{1} \,d\tau=\int_{\bar x^{0}}^{x^{0}}v \,dx^{0}\nonumber\\
&=&x^{0}-\bar x^{0}-\frac{1}{2}H\left((x^{0})^{2}-(\bar x^{0})^{2}\right)\,,\label{eq:dSworldline}
\eea
where $\bar x^{\mu}=x^{\mu}(\tau=0)$.
By using the mass-shell constraint we can also compute the evolution of the energy-momentum coordinates along the worldline. From the Hamilton equations we see that the spatial momentum coordinate $p_{1}$ is a constant of motion\footnote{However, the physical momentum $p^{1}$ is not.}. Then, using eq. (\ref{eq:dSmasslessconstraint})
\be
p_{0}-\bar p_{0}=p_{1}H (x^{0}- \bar x^{0})\,,\label{eq:dSEnergyEvolution}
\ee
where $\bar p_{0}=p_{0}(\tau=0)$.
Using this and the action of translations on momenta, last two lines of eq.~(\ref{eq:dStranslations}), one can compute the redshift of a particle measured by two distant observers, who compare the energy of a photon emitted by the first observer, Alice, and detected by the second one, Bob. 
In order to compute the redshift, we want to compare the energy measured by Alice in the origin of her reference frame\footnote{The superscript (or subscript) $X@Y$ indicates that  quantity is measure by observer X in the spatial origin of observer Y.}, $p_{0}^{A@A}$, with the energy measured by Bob in the origin of his reference frame, $p_{0}^{B@B}$. In order to do this, we first look at the evolution of the energy from the point of view of Alice. She will write eq.~(\ref{eq:dSEnergyEvolution}) as\footnote{Superscript or subscript $X$ indicates the value of a quantity at a generic point in space as inferred by observer X.}
\be
p_{0}^{A}- p_{0}^{A@A}=p_{1}^{A}H x^{0}_{A}=- p_{0}^{A@A}H x^{0}_{A}\,,\label{eq:dSAlicemassshell}
\ee
where we have taken into account the fact that the photon is emitted at Alice's origin ($x^{0}_{A@A}=0$) and that spatial momentum $p_{1}$ is constant along the worldline, and in particular the mass-shell constraint (\ref{eq:dSmasslessconstraint}) as applied at Alice's origin sets it to $p_{1}^{A}=-p_{0}^{A@A}$. When the photon is at Bob's origin, Alice will infer, using eq. (\ref{eq:dSAlicemassshell}):
\be
p_{0}^{A@B}=p_{0}^{A@A}(1-H x^{0}_{A@B})\,.
\ee
The energy measured by Bob when the photon crosses his spatial origin can be simply found by applying a translation to the above expression:
\be
p_{0}^{B@B}= \mathcal T_{\{a^{0},a^{1}\}} \triangleright  p_{0}^{A@B} = p_{0}^{A@B} = p_{0}^{A@A}(1-H x^{0}_{A@B})\,. \label{eq:dSenergycomparison}
\ee
So the redshift between Alice and Bob is
\be
z\equiv \frac{p_{0}^{A@A}-p_{0}^{B@B}}{p_{0}^{B@B}}= H x^{0}_{A@B} = H a^{0}\,,\label{eq:dSredshift}
\ee
where $a^{0}$ is the time translation parameter connecting Alice to Bob.

\section{de Sitter momentum space}
\label{sec:kP}

In this section we consider a situation that is somewhat complementary to the one we dealt with in the previous section. In fact, we consider a particle moving on flat (Minkowskian) spacetime and characterized by a curved momentum space, with de Sitter geometry. The radius of this de Sitter momentum space is given by the quantum deformation parameter $\ell$, which plays a role analogous to the one that usually $H$ has in de Sitter spacetime \cite{Amelino-Camelia:2013uya}. We are going to use a coordinatization of the de Sitter momentum space which is the analogous of the comoving coordinates we used for the de Sitter spacetime case of the previous section. In particular, the algebra of symmetries is the $\kappa$-Poincar\'e Hopf algebra, in the bicrossproduct basis \cite{Majid:1994cy,Lukierski:1993wxa}. In general a Hopf algebra  is not only defined by the commutation rules of its generators, but also by some additional structure, such as ``coproducts'' (fixing the action of generators on interacting particles and conservation rules in interactions) and ``antipodes''. These elements are not relevant for the kinematics of free particles, so here we will not deal with them.  We are reporting here results that are already known in the literature (see e.g. \cite{Amelino-Camelia:2013uya}) and we work at first order in $\ell$. All-order results can be found in \cite{Amelino-Camelia:2013uya}.

\subsection{$\kappa$-Poincar\'e Hopf algebra}

Similarly to the case of the previous section, describing momentum space as a maximally symmetric manifold  guarantees that the algebra of symmetries of the system has three symmetry generators in $1+1$ dimensions. We label them again $\{\mathcal P_{0},\mathcal P_{1}, \mathcal N\}$, and again they are generalizations of, respectively, the time translation, space translation and boost in Minkowski spacetime.
The algebra of these generators reads, at first order in $\ell$:
\begin{eqnarray}
\left\{\mathcal P_{0},\mathcal P_{1}\right\}&= & 0\,,\nonumber\\
\left\{\mathcal P_{0},\mathcal{N}\right\}&= &\mathcal P_{1} \,,\label{eq:Kalgebra}\\
\left\{\mathcal P_{1},\mathcal{N}\right\}&= &\mathcal P_{0}-\ell\left(\mathcal P_{0}^{2}+\frac{1}{2}\mathcal P_{1}^{2} \right)\,,\nonumber
\end{eqnarray}
and its Casimir is:
\be
\mathcal C_{\ell}= \mathcal P_{0}^{2}-\mathcal P_{1}^{2}-\ell \,\mathcal P_{0}\mathcal P_{1}^{2}\,.
\ee

Upon introducing the standard symplectic structure on the phase space coordinates $x^{\mu}$ and $p_{\mu}$, eq. (\ref{eq:sympl}), we can represent the generators as:
\bea
\mathcal P_{0}&=& p_{0}\,,\nonumber\\
\mathcal P_{1}&=& p_{1}\,,\label{eq:Kgeneratos}\\
\mathcal N&=& p_{1}x^{0}+p_{0}x^{1}-  \ell\left(x^{1}(p_{0})^{2}+\frac{x^{1}(p_{1})^{2}}{2}\right)\,,\nonumber
\eea
and the Casimir as:
\be
\mathcal C_{\ell}=p_{0}^{2}-p_{1}^{2}-\ell \,p_{0}\,p_{1}^{2}\,.\label{eq:KCasimir}
\ee

Since translation generators are represented trivially on momentum space coordinates, the action of translations on the phase space coordinates is the same as in flat spacetime and flat momentum space:
\bea
x^{0}_{B}&\equiv& \mathcal T_{\{a^{0},a^{1}\}} \triangleright x^{0}_{A} = x^{0}_{A}-a^{0}\,, \nonumber\\
x^{1}_{B}&\equiv& \mathcal T_{\{a^{0},a^{1}\}} \triangleright x^{1}_{A} = x^{1}_{A}-a^{1} \,,\label{eq:Ktranslations}\\
p^{B}_{0}&\equiv& \mathcal T_{\{a^{0},a^{1}\}} \triangleright p^{A}_{0} = p^{A}_{0}\,,\nonumber\\
p^{B}_{1}&\equiv& \mathcal T_{\{a^{0},a^{1}\}} \triangleright p^{A}_{1} = p^{A}_{1}\,,\nonumber
\eea
where, as in the previous section, the indices $A,B$ indicate two observers linked by a spacetime translation transformation with translation parameters $\{a^{0}, a^{1}\}$ and we used the general prescription of eq. (\ref{eq:FiniteTranslations}).

\subsection{Kinematics of massless particles with $\kappa$-Poincar\'e symmetries}

We use again the Hamiltonian formalism, with the Casimir (\ref{eq:KCasimir}) as Hamiltonian, in order to work out the evolution of a free massless particle's phase space coordinates:
\bea
\dot{x}^{0}&\equiv&\{\mathcal{C}_{\ell},x^{0}\}=
 2\,p_{0}-\ell \,p_{1}^{2}\,,\nonumber\\
\dot{x}^{1}&\equiv&\{\mathcal{C}_{\ell},x^{1}\}= -2\,p_{1}(1+\ell\, p_{0})\,,\label{eq:KHamilton}\\
\dot{p}_{0}&\equiv&\{\mathcal{C}_{\ell},{p}_{0}\}=   0\,,\nonumber\\
\dot{p}_{1}&\equiv&\{\mathcal{C}_{\ell},{p}_{1}\}= 0\,.\nonumber
\eea
Over-dots indicate derivatives with respect to the worldline's affine parameter $\tau$. Energy and spatial momentum satisfy the mass-shell constraint:
\be
p_{0}=-p_{1}(1-\frac{1}{2}\ell\, p_{1})\,.\label{eq:Kmasslessconstraint}
\ee
Since they are both constants of motion, there is no redshift in this model. However, the particle's worldline is deformed with respect to the standard Minkowskian one. In fact the coordinate velocity is:
\be
v\equiv \frac{\dot{x}^{1}}{\dot{x}^{0}}=1-\ell\, p_{1}\,,
\ee
and so the particle's worldline reads:
\be
x^{1}-\bar x^{1}\equiv\int_{0}^{\tau}\dot x^{1} \,d\tau=\int_{\bar x^{0}}^{x^{0}}v \,dx^{0}=\left(x^{0}-\bar x^{0}\right)\left(1-\ell p_{1}\right)\,,\label{eq:Kworldline}
\ee
where $\bar x^{\mu}=x^{\mu}(\tau=0)$. In oder to verify that this is not just a coordinates artefact, we need to make a proper relativistic analysis, comparing observations made by two observers, one at the emission and the other at the detection of the particle. In fact, it is now understood that in models with non-trivial momentum space geometry only observations made by local observers are reliable (as opposed to inferences made by distant observers) \cite{AmelinoCamelia:2010qv, AmelinoCamelia:2011bm}. We are going to show that indeed in this model the momentum space curvature leads to an effect that is complementary to the redshift characterizing the de Sitter spacetime case. This effect was dubbed ``lateshift'' in \cite{Amelino-Camelia:2013uya}, and amounts to an energy dependence of the arrival time of photons emitted simultaneously by one observer and detected by another far-away observer.

In order to compute the lateshift, we compare the times of arrival at the observer Bob of two photons emitted simultaneously by Alice in the origin of her reference frame with different energies\footnote{As for the previous section, the superscript (or subscript) $X@Y$ indicates that  quantity is measure by observer X in the spatial origin of observer Y. The Superscript or subscript $X$ indicates the value of a quantity at a generic point in space as inferred by observer X.}, $p_{0}^{A@A}$ and $\tilde p_{0}^{A@A}$. We assume that Bob detects in his spacetime origin the first photon (the one emitted from Alice with energy $p_{0}^{A@A}$). Then Bob is connected to Alice by a spacetime translation, with translation parameters $a^{0}$ and $a^{1}$ fixed by this condition.

The worldlines of the two photons  are inferred by Alice to be (we are using eq. (\ref{eq:Kworldline}) with $\bar x^{0}=0$ and $\bar x^{1}=0$):
\bea
x^{1}_{A}&=&x^{0}_{A}\left(1-\ell p_{1}^{A}\right)=x^{0}_{A}\left(1+\ell p_{0}^{A}\right)\,,\nonumber\\
\tilde x^{1}_{A}&=&\tilde x^{0}_{A}\left(1-\ell \tilde p_{1}^{A}\right)=\tilde x^{0}_{A}\left(1+\ell \tilde p_{0}^{A}\right)\,,
\eea
where we used the fact that we are working at the first order in $\ell$ and that energy and momentum are constants of motion along the worldline, so we can just write $p_{0}^{A}$ instead of $p_{0}^{A@A}$.
In order to write the worldlines as seen by the observer Bob, we use the translation transformations (\ref{eq:Ktranslations}):
\bea
x^{1}_{B}+a^{1}&=&(x^{0}_{B}+a^{0})\left(1+\ell p_{0}^{B}\right)\,,\nonumber\\
\tilde x^{1}_{B}+a^{1}&=&(\tilde x^{0}_{B}+a^{0})\left(1+\ell \tilde p_{0}^{B}\right)\,.
\eea
Since we ask that $x^{1}_{B}(x^{0}_{B}=0)=0$, we have to fix $a^{1}=a^{0}\left(1+\ell p_{0}^{B}\right)$. This defines the family of observers which detect the photon with energy $p_{0}^{B}$ in their spacetime origin and for whom the photons' worldlines read:
\bea
x^{1}_{B}&=&x^{0}_{B}\left(1+\ell p_{0}^{B}\right)\,,\nonumber\\
\tilde x^{1}_{B}&=&\tilde x^{0}_{B}\left(1+\ell \tilde p_{0}^{B}\right)+\ell a^{0} (\tilde p_{0}^{B}-p_{0}^{B})\,.
\eea
Then the photon with energy $\tilde p_{0}^{B}$ crosses Bob's spatial origin ($\tilde x^{1}_{B}=0$) at time:
\be
\tilde x^{0}_{B} = \ell a^{0} ( p_{0}^{B}-\tilde p_{0}^{B})\,. \label{eq:Klateshift}
\ee
From the point of view of Alice, the first particle reaches Bob at time:
\be
x^{0}_{A@B}= x^{0}_{B} +a^{0}=a^{0}\,,
\ee
while the second particle reaches Bob at time:
\bea
\tilde x^{0}_{A@B}&=&\tilde x^{0}_{B} +a^{0}=  a^{0} \left(1+\ell (p_{0}^{B}-\tilde p_{0}^{B})\right) \nonumber\\
&=& x^{0}_{A@B} \left(1+\ell (p_{0}^{A}-\tilde p_{0}^{A})\right)\,.
\eea
This formula for the lateshift resembles very closely the one we derived for the redshift in de Sitter spacetime, eq. (\ref{eq:dSenergycomparison}) and points out at the duality between kinematics in de Sitter spacetime and de Sitter momentum space which was discussed in detail in \cite{Amelino-Camelia:2013uya}.

\section{$q$-de Sitter phase space}
\label{sec:qdS}
The model we are going to focus on here provides an optimal setup to investigate the interplay between the effects of curvature in spacetime and in momentum space. In fact we study here the kinematics of free particles whose symmetries are described by the $q$-de Sitter Hopf algebra, \cite{Marciano:2010gq,Ballesteros:2004eu,Lukierski:1991pn} which is a quantum deformation of the de Sitter algebra. In particular, as discussed below, we choose the relation between the quantum deformation parameter and the Planck length $\ell$ and the inverse of the de Sitter radius $H$ such that in the $\ell\rightarrow 0$ limit the algebra contracts to the standard de Sitter algebra, which was discussed in section \ref{sec:dS}, while when $H\rightarrow 0$ the algebra contracts to the $\kappa$-Poincar\'e algebra, which was discussed in section \ref{sec:kP}. So a particle whose relativistic symmetries are the ones of the $q$-de Sitter algebra can be thought as moving in a phase space where both the spacetime side and the momentum space side are curved \footnote{In \cite{Barcaroli:2015xda} it was proposed a geometrical approach able to describe this kind of situations where the phase space can not be separated into a flat part and a curved part, and in particular the $q$-de Sitter case was studied as an example. Here we adopt the same phenomenological approach that we used in the previous sections, and we will focus on studying the kinematical properties of the model. }. We thus  expect the emergence of the phenomena of redshift and lateshift in the appropriate limits, while we also expect that in the general case, $\ell\neq 0$ and $H\neq 0$, the two effects are entwined. 
As explained also in the previous section concerning the $\kappa$-Poincar\'e Hopf algebra, the $q$-de Sitter Hopf algebra is characterized by additional structures  besides the commutation rules between the symmetry generators, such as the coproducts of the symmetries generators. Below we will  write down these as well for completeness, since the $q$-de Sitter algebra is far less known than the $\kappa$-Poincar\'e one. However, for the scopes of our analysis, which focuses on the kinematics of free particles, only the algebra of generators is relevant.

\subsection{The $q-$de Sitter Hopf algebra}

The $q-$de Sitter Hopf algebra has three generators in $1+1$ dimensions and is characterized by a  dimensionless quantum deformation parameter $w$, such that in the $w\rightarrow 0$ limit the algebra contracts to the standard de Sitter algebra. We will adopt a choice of basis for the algebra which contracts to the de Sitter algebra in the ``comoving'' basis used in section \ref{sec:dS}. The algebra of generators reads,  at all orders in the deformation parameter $w$ \cite{Marciano:2010gq}:
\begin{eqnarray}
{}\{\mathcal P_{0},\mathcal P_{1}\}&= & H\,\mathcal  P_{1}\,,\nonumber\\
{}\{\mathcal P_{0},\mathcal{N}\}&= & \mathcal P_{1}-H\,\mathcal{N}\,,\nonumber\\
{}\{\mathcal P_{1},\mathcal{N}\}&= &\cosh(w/2)\frac{1-e^{-2\frac{w\, \mathcal P_{0}}{H}}}{2w/H}\nonumber\\
&&-\frac{1}{H}\sinh(w/2)e^{-\frac{w\, \mathcal P_{0}}{H}}\Theta\,,
\end{eqnarray}
with
\[
\Theta=e^{\frac{w\, \mathcal P_{0}}{H}}(\mathcal P_{1}-H\,\mathcal{N})^{2}-H^{2}e^{\frac{w\, \mathcal P_{0}}{H}}\mathcal{N}^{2}.
\]
The coalgebra, which is associated to the action of symmetry transformations over interacting particles and to conservation rules in interactions, is:
\begin{eqnarray}
\Delta (\mathcal P_{0})&= & \mathcal P_{0}\otimes\mathbb{I}+\mathbb{I}\otimes \mathcal P_{0}\,,\nonumber\\
\Delta (\mathcal P_{1})&= & \mathcal P_{1}\otimes\mathbb{I}+e^{-w\frac{\mathcal P_{0}}{H}}\otimes \mathcal P_{1}\,,\\
\Delta(\mathcal{N})&= & \mathcal{N}\otimes\mathbb{I}+e^{-w\frac{\mathcal P_{0}}{H}}\otimes\mathcal{N}\,.\nonumber
\end{eqnarray}
The antipodes are the following:
\begin{eqnarray}
S (\mathcal P_{0})&= & -\mathcal P_{0}\,,\nonumber\\
S (\mathcal P_{1})&= & -e^{w\frac{\mathcal P_{0}}{H}}\mathcal P_{1}\,,\\
S (\mathcal{N})&= & -e^{w\frac{\mathcal P_{0}}{H}}\mathcal{N}\,,\nonumber
\end{eqnarray}
and, finally, the Casimir is
\bea
\mathcal{C}&=&H^{2}\frac{\cosh(w/2)}{w^{2}/4}\sinh^{2}\left(\frac{w\, \mathcal P_{0}}{2H}\right)-\frac{\sinh(w/2)}{w/2}\Theta\,.\,
\eea

As was already mentioned, $w$ is typically assumed to be a dimensionless combination of the two relevant scales of the model, the Planck length $\ell$ and the de Sitter radius $H^{-1}$ \cite{Marciano:2010gq,AmelinoCamelia:2003xp} . In particular, when $w= H \ell$, the $H\rightarrow 0$ contraction of the $q$-de Sitter algebra gives the $\kappa$-Poincar\'e algebra and the $\ell\rightarrow 0$ contraction gives the de Sitter algebra. 
We are interested exactly in this option for $w$, since we want to compare this model, with curvature on both momentum space side and spacetime side, with the models with curvature on only one of the two sides of the phase space, which were discussed in the previous sections. Therefore, we choose a basis for the $q$-de Sitter algebra such that, in the appropriate limits, one recovers the de Sitter algebra and the $\kappa$-Poincar\'e algebra in the bases used in the previous sections.

Setting $w=H \ell$, we will study the phenomenological properties of particles with $q$-de Sitter symmetries only up to the first order in $\ell$, $H$, and $\ell H$.
At this level of approximation the $q$-de Sitter algebra reads
\begin{eqnarray}
\{\mathcal P_{0},\mathcal P_{1}\}&= & H\,\mathcal  P_{1}\,,\nonumber\\
{}\{\mathcal P_{0},\mathcal{N}\}&= & \mathcal P_{1}-H\,\mathcal{N}\,,\\
{}\{\mathcal P_{1},\mathcal{N}\}&= &\mathcal  P_{0}-\ell\,\left(\mathcal P_{0}^{2}+\frac{\mathcal P_{1}^{2}}{2}\right)+\ell\, H\,\mathcal{N}\, \mathcal P_{1}\,,\nonumber
\end{eqnarray}
and has Casimir 
\begin{equation}
\mathcal{C}_{qdS}=\mathcal P_{0}^{2}-\mathcal P_{1}^{2}-\ell\,\mathcal  P_{0}\mathcal P_{1}^{2} + 2H\,\mathcal{N}\, \mathcal P_{1}+2\ell H\,\mathcal{N}\mathcal P_{0}\mathcal P_{1}\,.
\end{equation}

Similarly to what was done in the previous sections, we represent the $q$-de Sitter algebra on a phase space manifold with ordinary symplectic structure, eq. (\ref{eq:sympl}):
\begin{eqnarray}
\mathcal P_{0}&= & p_{0}-Hx^{1}p_{1}\,,\nonumber\\
\mathcal P_{1}&= & p_{1}\,,\label{eq:qdSgenerators}\\
\mathcal{N}&= & p_{1}x^{0}+p_{0}x^{1}-H\left(p_{1}(x^{0})^{2}+\frac{p_{1}(x^{1})^{2}}{2}\right)-\nonumber\\
&&-  \ell x^{1}\left((p_{0})^{2}+\frac{(p_{1})^{2}}{2}\right)+H\ell p_{1} x^{1}\left(p_{1}x^{0}+\frac{3}{2}p_{0}x^{1}\right)\nonumber\,.
\end{eqnarray}
The representation of the Casimir is
\begin{equation}
\mathcal{C}_{qdS}=p_{0}^{2}-p_{1}^{2}-\ell p_{0}p_{1}^{2}+2Hp_{1}^{2}x^{0}+2\ell H\, p_{0}p_{1}^{2}x^{0}\,.
\label{eq:qdSCasimir}
\end{equation}

The action of finite translations on the phase space coordinates is found by using the prescription of eq. (\ref{eq:FiniteTranslations}):
\bea
x^{0}_{B}&\equiv& \mathcal T_{\{a^{0},a^{1}\}} \triangleright x^{0}_{A} = x^{0}_{A}-a^{0}\,, \nonumber\\
x^{1}_{B}&\equiv& \mathcal T_{\{a^{0},a^{1}\}} \triangleright x^{1}_{A} = x^{1}_{A}(1+H a^{0})-a^{1}(1+\frac{1}{2} Ha^{0})\,, \nonumber\\
p^{B}_{0}&\equiv& \mathcal T_{\{a^{0},a^{1}\}} \triangleright p^{A}_{0} = p^{A}_{0}\,,\label{eq:qdStranslations}\\
p^{B}_{1}&\equiv& \mathcal T_{\{a^{0},a^{1}\}} \triangleright p^{A}_{1} = p^{A}_{1}(1-H a^{0})\,.\nonumber
\eea
This turns out to be the same as in standard de Sitter spacetime, eq. (\ref{eq:dStranslations}).

Before going on with our analysis, we note here that there has been a previous analysis \cite{AmelinoCamelia:2012it, Rosati:2015pga} of the phenomenological properties of particles living on a de Sitter spacetime and having a momentum space with de Sitter curvature. The model was built via a bottom-up approach, by deforming the de Sitter Casimir with a selection of $\ell$-dependent corrections and then working out the compatible algebra of symmetries. While the model considered there is indeed relativistic at the kinematical level, it is not clearly related to any quantum algebra of symmetries, as is the case here. In particular, there is no choice of the free parameters of the model in \cite{AmelinoCamelia:2012it, Rosati:2015pga} that can reproduce the $q$-de Sitter algebra and Casimir that we use here.

\subsection{Kinematics of massless particles with $q$-de Sitter phase space}\label{subs:dsSKinematics}

In order to derive the worldlines and momenta conservation laws of a free massless particle with q-de Sitter symmetries we adopt again a Hamiltonian procedure, using the $q$-de Sitter Casimir, eq. (\ref{eq:qdSCasimir}), as Hamiltonian. 
The variation of the phase space coordinates $\{x^{\mu},p_{\nu}\}$ with respect to the affine parameter $\tau$ is given by: 
\begin{eqnarray}
\dot{x}^{0}&\equiv&\{\mathcal{C}_{qdS},x^{0}\}=
 2p_{0}-\ell p_{1}^{2}(1-2H\, x^{0})\,,\nonumber\\
\dot{x}^{1}&\equiv&\{\mathcal{C}_{qdS},x^{1}\}=
 -2p_{1}(1+\ell\, p_{0})(1-2H\, x^{0})\,,\label{eq:qdSHamilton}\\
\dot{p}_{0}&\equiv&\{\mathcal{C}_{qdS},{p}_{0}\}=  
 -2H\, p_{1}^{2}(1+\ell\, p_{0})\,,\nonumber\\
\dot{p}_{1}&\equiv&\{\mathcal{C}_{qdS},{p}_{1}\}= 0\,.\nonumber
\end{eqnarray}
Because of spacetime curvature, the momenta are not constants of motion, in analogy to the de Sitter spacetime case. However, the evolution of momenta along the particle's worldline now depends on the Planck length $\ell$ besides the de Sitter radius $H$.
Again, one can check that in the $\ell=0$ limit one recovers the standard de Sitter worldlines and conservation laws for momenta, eq. (\ref{eq:dSHamilton}), while in the $H=0$ limit one recovers the $\kappa$-Poincar\'e relations, eq. (\ref{eq:KHamilton}). These worldlines were also derived in \cite{Barcaroli:2015xda}, where the focus was however on building a phase space geometrical picture rather then working out the phenomenology for this model \footnote{In  \cite{Barcaroli:2015xda} different conventions are used for the symplectic structure on phase space, so that the worldlines written there are a mapping of these ones.}. 
%

The massless on-shell relation, $\mathcal C_{qdS}=0$, fixes the relation between $p_{0}, p_{1}$ and $x^{0}$:
 \begin{equation}
p_{0} = -p_{1}\left(1-H x^{0}-\ell p_{1} \left(\frac{1}{2}-H x^{0}\right)\right)\,,\label{eq:qdSmasslessonshellness}
 \end{equation}
 where $p_{1}$ is a constant of motion and the overall sign was chosen so to have positive coordinate velocity:
 \be
 v\equiv\frac{\dot x^{1}}{\dot x^{0}}=1-H x^{0}-\ell p_{1}(1-2 H x^{0})\,.
 \ee
 The particle's worldline is then found to be:
 \bea
&&x^{1}-\bar x^{1}\equiv\int_{0}^{\tau}\dot x^{1} \,d\tau=\int_{\bar x^{0}}^{x^{0}}v \,dx^{0}\nonumber\\
&&=\left(x^{0}-\bar x^{0}\right)\left(1-\ell p_{1}\right)-\frac{1}{2} H \left((x^{0})^{2}-(\bar x^{0})^{2}\right)\left(1-2 \ell p_{1} \right)\,,\nonumber\\\label{eq:qdSworldline}
\eea
where $\bar x^{0}=x^{0}(\tau=0)$ and $\bar x^{1}=x^{1}(\tau=0)$. This worldline can also be written in terms of the initial energy measured by the observer Alice whose spacetime origin coincides with the emission of the particle. In fact, at the emission $x^{0}=0$ and so the mass-shell constraint simplifies to:
\begin{equation}
\mathcal{C}_{qdS}(x^0=0) = 0 \Rightarrow {p}_1 = - {p}_0^{A@A}\left(1  -\frac{\ell}{2}{p}_0^{A@A} \right)\,,\label{eq:qdSmassshellatemission}
\end{equation} 
where we have used the notation introduced in the previous sections, such that the super- (or sub-) script $X@Y$ indicates that a quantity is measured by the observer X at the  origin of observer Y (and in particular $X@X$ stands for a measurement made by the observer X at her own spatial origin). The super- (or sub-) script $X$ indicates a quantity measured in the coordinate system of observer $X$ at a generic point. 
So the worldline as inferred by the observer Alice reads:
\bea
x^{1}_{A}=x^{0}_{A}\left(1+\ell p_{0}^{A@A}\right)\label{eq:qdSworldline2}-\frac{1}{2} H (x^{0}_{A})^{2}\left(1+2 \ell p_{0}^{A@A} \right)\,.\,\,
\eea
The non-linearity in the coordinate time is a sign of  curvature of spacetime, while the explicit energy-dependence is a  sign of curvature of momentum space.

In order to expose clearly the effects ascribed to spacetime curvature, the ones due to momentum space curvature and the ones due to the interplay between the two, we are going to look again at the redshift and the lateshift, adopting the same procedures used in the previous sections.

%
%
\subsection{Energy-dependent redshift}\label{subs:qdSredshift}

In this section we follow closely the analysis presented in the de Sitter spacetime case, where we computed the amount of redshift affecting a photon traveling between two distant observers, Alice (at emission) and Bob (at detection).
The evolution of the energy of the photon along its worldline as seen by the observer Alice is given by eqs. (\ref{eq:qdSmasslessonshellness}) and (\ref{eq:qdSmassshellatemission}):
\bea
p_{0}^{A}-p_{0}^{A@A}&=&  H p_{1}^{A} x^{0}_{A}\left(1- \ell p_{1}  \right) \nonumber\\
&=&  - H  {p}_0^{A@A}x^{0}_{A}\left(1+\frac{\ell}{2}{p}_0^{A@A} \right) \,.
\eea
Then when the photon is at Bob's origin, Alice will infer its energy to be:
\be
p_{0}^{A@B}=  p_{0}^{A@A}\left(1- H x^{0}_{A@B}\left(1+\frac{\ell}{2}{p}_0^{A@A} \right) \right)\,. \label{eq:qdSenergies1}
\ee

The energy measured by Bob at his spatial origin is obtained via a translation:
\be
p_{0}^{B@B}=\mathcal T_{a^{0},a^{1}}\triangleright p_{0}^{A@B} = p_{0}^{A@B}\,.  \label{eq:qdSenergies2}
\ee
So the redshift between Alice and Bob is:
\bea
z&\equiv& \frac{p_{0}^{A@A}-p_{0}^{B@B}}{p_{0}^{B@B}}= H x^{0}_{A@B}\left(1+\frac{\ell}{2}{p}_0^{A@A} \right) \nonumber\\&=& H a^{0}\left(1+\frac{\ell}{2}{p}_0^{A@A} \right)\,,
\eea
where $a^{0}$ is the time-translation parameter connecting Alice and Bob.
Notice that this formula for the redshift contains an energy-dependent correction term with respect to the on valid in the de Sitter spacetime case, eq. (\ref{eq:dSredshift}). This correction term can be seen as the effect of  the interplay between spacetime and momentum space curvature.

\subsection{$q$-de Sitter lateshift}\label{subs:qdSlateshift}

As seen in the section about de Sitter momentum space, the lateshift is a characteristic feature of momentum space curvature, which causes the travel time of massless particles to depend on their energy.
Here we want to explore this feature in the case where both spacetime and momentum space are curved, relying on the same line of reasoning followed in the previous section. So we compare the times of arrival at the observer Bob of two photons emitted simultaneously by Alice in the origin of her reference frame, but with different energies, $p_{0}^{A@A}$ and $\tilde p_{0}^{A@A}$ . Bob is defined to detect the first photon (the one emitted by Alice with energy $p_{0}^{A@A}$) in his spacetime origin. 
The worldlines of the two photons are inferred by Alice to be (see eq. (\ref{eq:qdSworldline2})):
\bea
x^{1}_{A}&=&x^{0}_{A}\left(1+\ell p_{0}^{A@A}\right)-\frac{1}{2} H (x^{0}_{A})^{2}\left(1+2 \ell p_{0}^{A@A} \right)\,,\nonumber\\
\tilde x^{1}_{A}&=&\tilde x^{0}_{A}\left(1+\ell \tilde p_{0}^{A@A}\right)-\frac{1}{2} H (\tilde x^{0}_{A})^{2}\left(1+2 \ell \tilde p_{0}^{A@A} \right)\,.\,\,
\eea
By using the translation transformations (\ref{eq:qdStranslations}) we can deduce the worldlines as seen by the observer Bob:
\bea
x^{1}_{B}(1-H a^{0})&=&a^{1}(\frac{1}{2}H a^{0}-1)+(x^{0}_{B}+a^{0})\left(1+\ell p_{0}^{A@A}\right)\nonumber\\&&-\frac{1}{2} H (x^{0}_{B}+a^{0})^{2}\left(1+2 \ell p_{0}^{A@A} \right)\,,\nonumber\\
\tilde x^{1}_{B}(1-H a^{0})&=&a^{1}(\frac{1}{2}H a^{0}-1)+(\tilde x^{0}_{B}+a^{0})\left(1+\ell \tilde p_{0}^{A@A}\right)\nonumber\\
&&-\frac{1}{2} H (\tilde x^{0}_{B}+a^{0})^{2}\left(1+2 \ell \tilde p_{0}^{A@A} \right)\,.
\eea
The relation between the translation parameters $a^{0}$ and $a^{1}$ is fixed by the requirement that the photon with energy $p_{0}^{A@A}$ goes through the spacetime origin of Bob:
\bea
a^{1}&=&a^{0}+\ell a^{0 } p_{0}^{A@A}-\frac{1}{2}H \ell (a^{0})^{2}  p_{0}^{A@A}\,.
\eea
So  the worldlines can be written as:
\bea
x^{1}_{B}&=&x^{0}_{B}\left(1+\ell p_{0}^{A@A}\right)\nonumber\\
&&- H x^{0}_{B}\left(\frac{1}{2} x^{0}_{B} +\ell p_{0}^{A@A}\left(x^{0}_{B}+a^{0}\right)\right)\,,\nonumber\\
\tilde x^{1}_{B}&=&\tilde x^{0}_{B}\left(1+\ell \tilde p_{0}^{A@A}\right)+\ell a^{0}(\tilde p_{0}^{A@A}-p_{0}^{A@A} )\nonumber\\
&&- H \tilde x^{0}_{B} \left(\frac{1}{2} \tilde x^{0}_{B} +\ell \tilde p_{0}^{A@A}\left(\tilde  x^{0}_{B}+a^{0}\right)\right)\,.\nonumber\\
\eea
The worldline parameters $p_{0}^{A@A}$ and $\tilde p_{0}^{A@A}$ can be rewritten using  eqs. (\ref{eq:qdSenergies1})-(\ref{eq:qdSenergies2}) and observing that $x^{0}_{A@B}=x^{0}_{B@B}+a^{0}$ and $\tilde x^{0}_{A@B}=\tilde x^{0}_{B@B}+a^{0}$. Then one finds: 
\bea
\ell p_{0}^{A@A}&=&\ell p_{0}^{B@B}(1+H x^{0}_{A@B}) = \ell p_{0}^{B@B}(1+H a^{0}) \nonumber\\
\ell \tilde p_{0}^{A@A}&=&\ell \tilde p_{0}^{B@B}(1+H \tilde x^{0}_{A@B})=\ell \tilde p_{0}^{B@B}(1+H  a^{0})\nonumber \,,
\eea
where we have also used the fact that by definition of the observer Bob $x^{0}_{B@B}=0$ and $\tilde x^{0}_{B@B}=0+\mathcal O(\ell)$.

Then one can find the time $\tilde x^{0}_{B@B}$ at which the second photon intercepts Bob's spatial origin, in Bob's coordinate system:
\be
\tilde x^{0}_{B@B}=\ell a^{0}\left[  (p_{0}^{B@B} -\tilde p_{0}^{B@B})(1+H a^{0}) \right]\,.
\ee
This equation is analogous to the one found in the de Sitter momentum space case, eq. (\ref{eq:Klateshift}), with a correction term depending on spacetime curvature via the inverse of the de Sitter radius $H$. 
As done in that case, we can also compute the delay inferred by Alice:
\bea
\tilde x^{0}_{A@B} &=& \tilde x^{0}_{B@B}+a^{0}\nonumber \\
&=& a^{0} \left[1+ \ell \left( (p_{0}^{B@B} -\tilde p_{0}^{B@B})(1+H a^{0}) \right)\right]\nonumber\\
&=&  x^{0}_{A@B} \left[1+ \ell \left( (p_{0}^{B@B} -\tilde p_{0}^{B@B})(1+H a^{0}) \right)\right]\nonumber\\
&=&  x^{0}_{A@B} \left[1+ \ell \left( p_{0}^{A@A} -\tilde p_{0}^{A@A}\right)\right]\,.
\eea
In this model, contrary to what happens in the flat-spacetime case of $\kappa$-Poincar\'e, Alice and Bob do not write formally identical equations for the delay in their own reference frame, because of the effects of energy redshift. However, the actual value of the delay they would get from their formulas is the same. 

\section{Conclusions}
The study of relativistic deformations of particles' kinematics on curved spacetimes is of fundamental importance for the purposes of quantum gravity phenomenology, as this mostly deals with propagation of particles over cosmological distances. We have here investigated the kinematical predictions of a quantum deformation of the de Sitter algebra, known as $q$-de Sitter. This Hopf algebra describes the symmetries of a particle living on a de Sitter spacetime and having a curved manifold of momenta, with de Sitter geometry. Using a Hopf algebra guarantees that besides the deformation of the mass-shell condition, also other ingredients of relativistic theories, such as conservation laws in interactions, can be coherently introduced. The study we present here is indeed the first one to derive in detail the kinematical predictions of models with Hopf algebraic symmetries underlying a curved spacetime. We find that a particle with $q$-de Sitter symmetries is subject to both redshift of its energy during propagation and the so-called lateshift, \emph{i.e.} an energy dependence of the time of travel. These are, respectively, well-known features of curved spacetime and curved momentum space models, and so their simultaneous presence signals that the model we are considering is indeed curved on both sides of the phase space. However, both of the effects are modified with respect to what one would have were the curvature to be on  one side of the phase space only. In particular, the amount of redshift depends not only on the travel time of the particle, but also on its energy. Moreover, the travel time of a particle between two observers depends on the spacetime curvature besides the energies themselves (and so it has a modified dependence on the distance between emission and detection with respect to the flat spacetime case).

\acknowledgements
GG acknowledges support  from the John Templeton Foundation.	

\bibliography{RelativeLocality}

\end{document}